\documentclass[trackchanges]{aastex701}
\usepackage{amsmath}
\usepackage{hyperref}
\begin{document}

\title{Modeling the Thermal Low-Frequency Radio Sun with Ray Tracing}

\author[0000-0001-6855-5799]{Peijin Zhang}
\affiliation{Center for Solar-Terrestrial Research, New Jersey Institute of Technology, Newark, NJ 07102, USA}
\email{}
\author[0000-0002-0660-3350]{Bin Chen}
\affiliation{Center for Solar-Terrestrial Research, New Jersey Institute of Technology, Newark, NJ 07102, USA}
\email{}
\author[0000-0002-0508-0332]{Gregory Fleishman}
\affiliation{Center for Solar-Terrestrial Research, New Jersey Institute of Technology, Newark, NJ 07102, USA}
\email{}
\author[0000-0001-8644-8372]{Alexey Kuznetsov}
\affiliation{Institute of Solar-Terrestrial Physics, Siberian Branch of the Russian Academy of Sciences, Irkutsk 664033, Russia}
\email{}
\author[0000-0003-1759-4354]{Cooper Downs}
\affiliation{Predictive Science Inc., San Diego, CA 92121, USA}
\email{}
\author[0000-0002-2325-5298]{Surajit Mondal}
\affiliation{National Centre for Radio Astrophysics, Tata Institute of Fundamental Research, Pune 411007, India}
\email{}
\author[0000-0003-2872-2614]{Sijie Yu}
\affiliation{Center for Solar-Terrestrial Research, New Jersey Institute of Technology, Newark, NJ 07102, USA}
\email{}

\begin{abstract}
% why
Incoherent radio emission at meter--decimeter wavelengths provides a key diagnostic of the coronal thermal plasma, but at frequencies below $\sim$\,1\,GHz coronal refraction can substantially bend ray paths and modify the apparent source size and brightness distribution. 
% how
We develop a forward-modeling framework that combines refractive ray tracing through a global 3D coronal model with radiative transfer along each ray. The method tracks the ray-tube cross-sectional area $S(s)$ using a step-wise perturbation retracing approach and incorporates a geometric magnification term proportional to $d\ln S/ds$ to enforce flux conservation under focusing/defocusing. Thermal free--free emission and absorption are then computed with the \texttt{GRFF} radiative transfer code to produce synthetic radio maps over 40--800\,MHz.
% results
Applying the framework to Carrington rotation 2298, we find that including propagation effects allows the quiet-Sun background spectrum to be well reproduced. However, active region brightness is less accurately modeled, suggesting that additional physical factors should be considered in future work.
% conclusion
These results establish a physics-based method for generating low-frequency quiet-Sun synthetic images suitable for quantitative comparison with interferometric observations and for assessing how propagation effects shape the observed morphology.
\end{abstract}

\section{Introduction}

Solar quiescent radio emission at meter and decimeter wavelengths provides a key diagnostic of the global coronal thermal plasma. At frequencies below $\sim1$\,GHz, the emission is dominated by thermal emission (free--free and gyro-resonance), and the observable brightness distribution is set by a combination of the 3D coronal density/temperature structure, magnetic field structure, which set up frequency-dependent optical depth.. In addition, strong gradients in the coronal refractive index bend radio rays, shifting the apparent radio emission generation location, altering limb brightness, and changing the apparent source size as a function of frequency (e.g., \citealt{1971A&A....10..362S,1985ARA&A..23..169D,2011SoPh..273..309S}).

The quiet-Sun brightness temperature spectrum and its frequency-dependent apparent size have been measured across a wide range of radio frequencies, from submillimeter \citep{2023A&A...670C...5A} and centimeter/microwave  \citep{1991ApJ...370..779Z} bands to meter wavelengths \citep[e.g.,][]{2004A&A...426..329S,2006ApJ...648..707R,2018SoPh..293...97M,2022ApJ...932...17Z}. Multifrequency imaging has been used to infer coronal electron density and temperature distributions \citep{2015A&A...583A.101M,2026ApJ...999..237M}. Such inferences require careful forward modeling because propagation effects become increasingly important at low frequencies \citep{1994ApJ...426..774B}, where refraction and scattering can bias the apparent solar radius, modify centre-to-limb profiles, and change the relationship between observed brightness and local plasma properties \citep{2020ApJ...903..126S}.

Forward modeling that connects a physics-based coronal model to synthetic radio images is therefore crucial for quantitative comparisons with modern low-frequency observations. The standard approach, that solves the radiative-transfer equation along straight lines of sight (LOS) through a 3D coronal volume, breaks down at $\lesssim 300$~MHz, where the geometric-optics rays deviate substantially from straight lines. Mapping between the image plane and coronal locations as well as the conservation of the flux under focusing/defocusing by refraction should properly be taken into account at these long wavelengths.

In this work we develop a refractive ray-tracing and transfer framework for the thermal emission from the Sun at 40--800~MHz. Our method explicitly traces rays through an inhomogeneous plasma and augments the radiative-transfer equation with a geometric term that accounts for changes in the ray-tube cross-sectional area $S(s)$. In the absence of emission and absorption, conservation of energy flow implies $I\,S=\mathrm{const}$; when $S$ varies, this leads to an additional term proportional to $d\ln S/ds$ in the transfer equation. We estimate $S(s)$ numerically using a step-wise perturbation retracing method, allowing us to apply a local magnification correction that is naturally consistent with flux conservation.
We demonstrate the framework using the PSI MAS global MHD corona as input and compute maps of the thermal emission from the Sun. 

The paper is organized as follows. Section~\ref{sec:ray_tube_scaling} introduces the ray-tube magnification formalism and the perturbation-based area estimate used during ray tracing. Section~3 describes the PSI MAS model setup and the end-to-end procedure used to produce synthetic images for a representative snapshot (\mbox{2025-06-08 20:00 UT}). We discuss implications for quiet-Sun imaging and future comparisons with interferometric data in Sections~4--5.

\section{Radiation transfer with local magnification factor}\label{sec:ray_tube_scaling}

Refraction in an inhomogeneous plasma bends neighboring rays differently, causing a narrow bundle of trajectories (a \emph{ray tube}) to either converge (focusing) or diverge (defocusing). This geometric lensing changes the ray-tube cross-sectional area \(S\) and therefore modifies the flux carried by the bundle. In the geometric-optics limit, and in the absence of absorption/emission, the energy flow through the tube is conserved, so that
\begin{equation}
I(s)\,S(s)=\mathrm{const},
\label{eq:IS_conserve}
\end{equation}
where \(I\) is the specific intensity (or ray-tube intensity proxy used by the transfer solver) and \(s\) is the path-length coordinate along the central ray. In other words, focusing increases \(I\) by decreasing \(S\), while divergence decreases \(I\) by increasing \(S\). In practice, we incorporate this effect via a step-wise estimate of the \emph{area magnification} of the ray tube during numerical ray tracing.

With emission and absorption included, conservation of the energy flow through the tube implies that the standard radiative-transfer equation acquires an additional geometric term associated with the varying source area. %Writing the transfer equation along the propagation coordinate \(s\),
\begin{equation}
\frac{dI}{ds}= j - \kappa I - \frac{I}{S}\frac{dS}{ds}
\label{eq:rt_with_magnification}
\end{equation}
where $s$ is the path-length coordinate along the central ray, \(j\) and \(\kappa\) are the emissivity and absorption coefficient, respectively. The quantity
\begin{equation}
\frac{1}{S}\frac{dS}{ds}=\frac{d\ln S}{ds}
\label{eq:local_magnification}
\end{equation}
is the \emph{local magnification factor}. If the ray-tube area does not change (\(dS/ds=0\)), this term is zero, and Eq.~\eqref{eq:rt_with_magnification} reduces to the usual 1D transfer equation.
A positive value of this term indicates defocusing, and a negative value of this term indicates focusing.

\begin{figure}[ht!]
\centering
\includegraphics[width=0.6\textwidth]{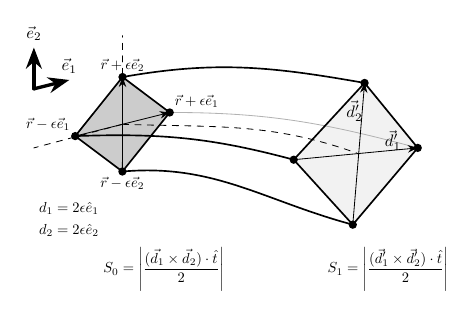}
\caption{Ray-tube geometry and transverse basis; perturbed rays probe the cross-sectional area change.}
\label{fig:tube}
\end{figure}

Figure~\ref{fig:tube} illustrates the ray-tube geometry. At each integration step, we define a local orthonormal basis \(\{\hat{\mathbf{t}},\hat{\mathbf{e}}_1,\hat{\mathbf{e}}_2\}\), where \(\hat{\mathbf{t}}\) is the propagation direction of the central ray and \(\hat{\mathbf{e}}_1,\hat{\mathbf{e}}_2\) span the plane normal to \(\hat{\mathbf{t}}\). Using \(\hat{\mathbf{e}}_1\) and \(\hat{\mathbf{e}}_2\), we construct two perturbed rays that sample the local deformation of the bundle, advance them by one step with the same integrator, and estimate the updated cross-sectional area from the resulting separation vectors. This procedure yields a local scaling factor that can be applied to the transfer equation through Eq.~\eqref{eq:IS_conserve}.

\subsection{Step-wise perturbation retracing algorithm}
We follow the unmagnetized Hamiltonian ray equations:
\begin{equation}
\frac{d\mathbf{r}}{dt}=\frac{\partial \omega}{\partial \mathbf{k}},\qquad
\frac{d\mathbf{k}}{dt}=-\frac{\partial \omega}{\partial \mathbf{r}},
\label{eq:hamilton_rays}
\end{equation}
where $\mathbf{r}(x,y,z)$ is the location of the photon, $\mathbf{k}(k_x, k_y, k_z)$ is the wave vector.
With dispersion \(\omega^2=\omega_{pe}(\mathbf{r})^2 + k^2\) in the code's normalized units. The central-ray state is a 6-dimension real-number array \(\mathbf{x}\equiv(\mathbf{r},\mathbf{k})\in\mathbb{R}^6\), advanced using an RK4 stepper with fixed \(\Delta t\).

At step \(i\), let \(\mathbf{r}_0^{i}\) be the central-ray position before the RK4 update and \(\mathbf{r}_0^{i+1}\) after the update. We define the local propagation unit vector
\begin{equation}
\hat{\mathbf{t}}^{i}=\frac{\mathbf{r}_0^{i+1}-\mathbf{r}_0^{i}}{\left\|\mathbf{r}_0^{i+1}-\mathbf{r}_0^{i}\right\|}.
\label{eq:t_hat}
\end{equation}
To form a stable transverse basis, we choose a reference axis \(\mathbf{a}\) that is least aligned with \(\hat{\mathbf{t}}^{i}\) (e.g., \(\mathbf{a}=\hat{\mathbf{z}}\) unless \(|\hat{t}_z| \approx 1\), in which case \(\mathbf{a}=\hat{\mathbf{y}}\)), and set
\begin{equation}
\hat{\mathbf{e}}_1^{i}=\frac{\mathbf{a}\times \hat{\mathbf{t}}^{i}}{\left\|\mathbf{a}\times \hat{\mathbf{t}}^{i}\right\|},
\qquad
\hat{\mathbf{e}}_2^{i}=\hat{\mathbf{t}}^{i}\times \hat{\mathbf{e}}_1^{i}.
\label{eq:e1e2}
\end{equation}
We then generate two perturbed initial conditions at the same step origin \(\mathbf{r}_0^{i}\):
\begin{equation}
\mathbf{r}_{1}^{i}=\mathbf{r}_0^{i} + \epsilon_i\,\hat{\mathbf{e}}_1^{i},
\qquad
\mathbf{r}_{2}^{i}=\mathbf{r}_0^{i} + \epsilon_i\,\hat{\mathbf{e}}_2^{i},
\label{eq:perturb_init}
\end{equation}
with \(\epsilon_i\) chosen proportional to the step length,
\begin{equation}
\epsilon_i = \eta\,\left\|\mathbf{r}_0^{i+1}-\mathbf{r}_0^{i}\right\|,
\label{eq:epsilon}
\end{equation}
where \(\eta\) is a user-defined perturbation ratio (parameter \texttt{perturb\_ratio} in the code). The perturbed rays share the same wavevector at step \(i\), \(\mathbf{k}_{1}^{i}=\mathbf{k}_{2}^{i}=\mathbf{k}_0^{i}\), and are advanced by one RK4 step to obtain \(\mathbf{r}_{1}^{i+1}\) and \(\mathbf{r}_{2}^{i+1}\).

The two transverse separation vectors at the end of the step are
\begin{equation}
\delta \mathbf{r}_1^{i+1}=\mathbf{r}_{1}^{i+1}-\mathbf{r}_0^{i+1},
\qquad
\delta \mathbf{r}_2^{i+1}=\mathbf{r}_{2}^{i+1}-\mathbf{r}_0^{i+1},
\label{eq:sep_vec}
\end{equation}
which defines a local parallelogram in the transverse plane. The corresponding ray-tube cross-sectional area is estimated by
\begin{equation}
S^{i+1}\approx \left|\frac{\left(\delta \mathbf{r}_1^{i+1}\times \delta \mathbf{r}_2^{i+1}\right)}{2}\cdot \hat{\mathbf{t}}^{i}\right|.
\label{eq:area_est}
\end{equation}
Finally, we define a dimensionless \emph{area scaling factor} (or magnification) for this step by normalizing out the imposed perturbation area \(\epsilon_i^2\):
\begin{equation}
\mathcal{M}^{i+1}\equiv \frac{S^{i+1}}{\epsilon_i^2}.
\label{eq:mag_factor}
\end{equation}
Values \(\mathcal{M}>1\) indicate divergence (tube expansion), while \(\mathcal{M}<1\) indicate focusing (tube contraction). In the absence of radiative sources/sinks, Eq.~\eqref{eq:IS_conserve} implies that the intensity update due to geometric lensing over one step is
\begin{equation}
I^{i+1}=I^{i}\,\frac{S^{i}}{S^{i+1}}
\ \ \propto\ \ 
\frac{1}{\mathcal{M}^{i+1}},
\label{eq:intensity_update}
\end{equation}
so that focusing increases \(I\) and defocusing decreases \(I\). In the full transfer calculation with emission/absorption, this factor can be incorporated either as a multiplicative correction per step or equivalently as an additive focusing term in the differential form of the transfer equation via \(d(\ln S)/ds\).

Figure~\ref{fig:tube} summarizes the construction: the central ray defines the local normal plane and transverse basis; two perturbed rays probe the deformation of the ray tube; the resulting area magnification \(\mathcal{M}\) provides the geometric scaling needed to conserve \(I\,S\) under refraction.

\section{Synthesizing radio images with ray tracing}

Our modeling involves two main steps: (1) refractive ray tracing through the inhomogeneous coronal plasma to obtain the propagation path and geometric magnification, and (2) solving the radiative-transfer equation along each ray (Equation~\ref{eq:rt_with_magnification}) to synthesize an image in the observer's plane.
Ray tracing requires a 3D coronal model that provides the electron density $N_e(\mathbf{r})$ to determine the refractive index and propagation path.
Solving the radiative-transfer equation requires all three parameters: $N_e(\mathbf{r})$, the electron temperature $T_e(\mathbf{r})$, and the magnetic field $\mathbf{B}(\mathbf{r})$.

\subsection{Coronal model}

\begin{figure}[h!]
    \centering
    \includegraphics[width=0.98\linewidth]{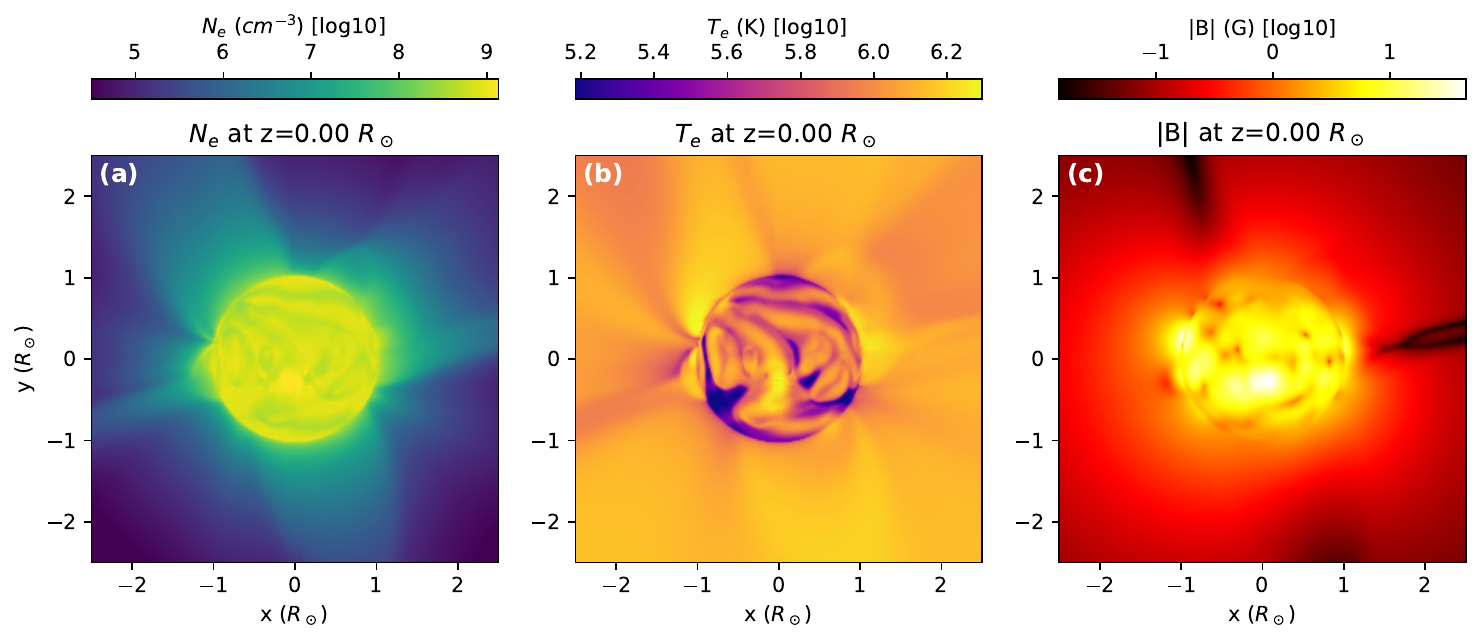}
    \caption{Plane of sky 2D slices through the PSI MAS model at $z=0$, showing the electron density $N_e$, electron temperature $T_e$, and magnetic-field strength $|\mathbf{B}|$ used for the ray-tracing calculations.}
    \label{fig:model_slice}
\end{figure}

\begin{figure}[h!]
    \centering
    \includegraphics[width=0.8\linewidth]{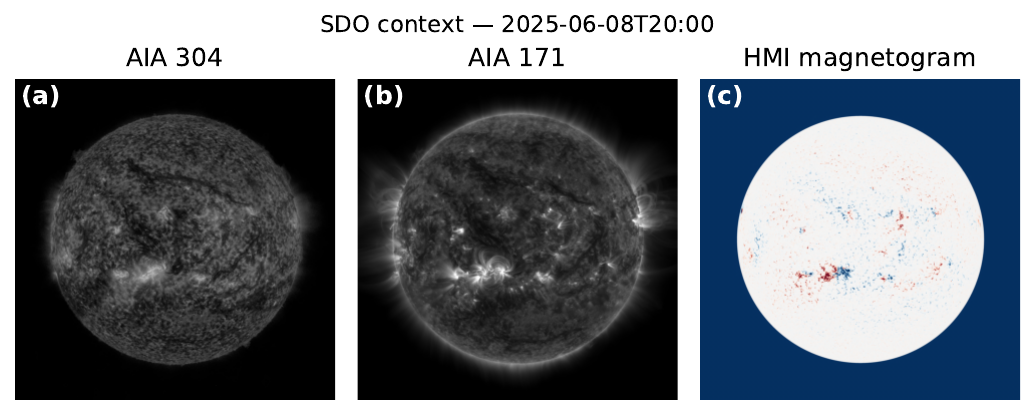}
    \caption{Context solar images from the Solar Dynamics Observatory (SDO).}
    \label{fig:sdo}
\end{figure}

Here we use the MAS (Magnetohydrodynamic Algorithm outside a Sphere) model developed by Predictive Science Inc. (PSI) \citep{1996AIPC..382..104M}.
PSI MAS is a global 3D, time-dependent MHD model of the solar corona and inner heliosphere: it solves the coupled plasma flow and magnetic-field equations in a spherical domain outside the Sun, using photospheric magnetic maps as boundary conditions to produce self-consistent distributions of density, temperature, velocity, and magnetic field. We use the coronal model MHD solution corresponding to CR2298 obtained from a standard series of MAS calculations developed by PSI for every Carrington Rotation\footnote{\href{https://www.predsci.com/mhdweb2/tools/data-access}{www.predsci.com/mhdweb2/tools/data-access}}. This model, labeled `HMI CR 2298 Thermo 2', utilizes empirical coronal heating functions similar to \citep{2009ApJ...690..902L}. It is run at a relatively modest global resolution, with a spherical grid size of $255\times143\times300$ in radius ($r$), co-latitude ($\theta$), and longitude ($\phi$). Figure~\ref{fig:model_slice} shows example plane of sky slice of $N_e$, $T_e$, and $|\mathbf{B}|$, rotated to Carrington longitude $140^\circ$ to align the model with the observing time \mbox{2025-06-08 20:00 UT}. Figure~\ref{fig:sdo} shows the corresponding context images from SDO at \mbox{2025-06-08 20:00:00 UT}.

\subsection{Ray-tracing}

\begin{figure}[ht!]
    \centering
    \includegraphics[width=0.6\linewidth]{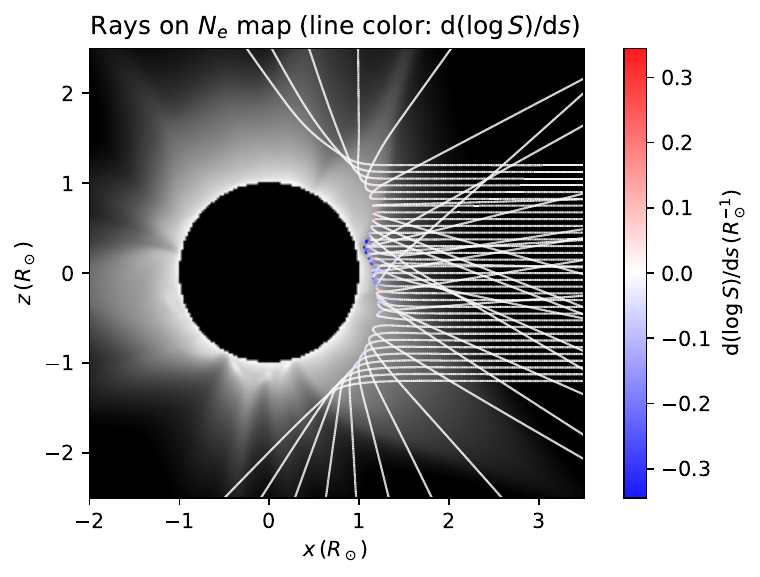}
    \caption{Ray-tracing trajectories overlaid on an electron-density map from the PSI MAS coronal model. The 32 rays are launched from $x=-3.5$, $y=0$, $z=(-1.2, 1.2)$, colors indicate the variation of the magnification factor, positive indicates defocusing, negative indicates focusing.}
    \label{fig:ray_on_ne}
\end{figure}

\begin{figure}
    \centering
    \includegraphics[width=0.63\linewidth]{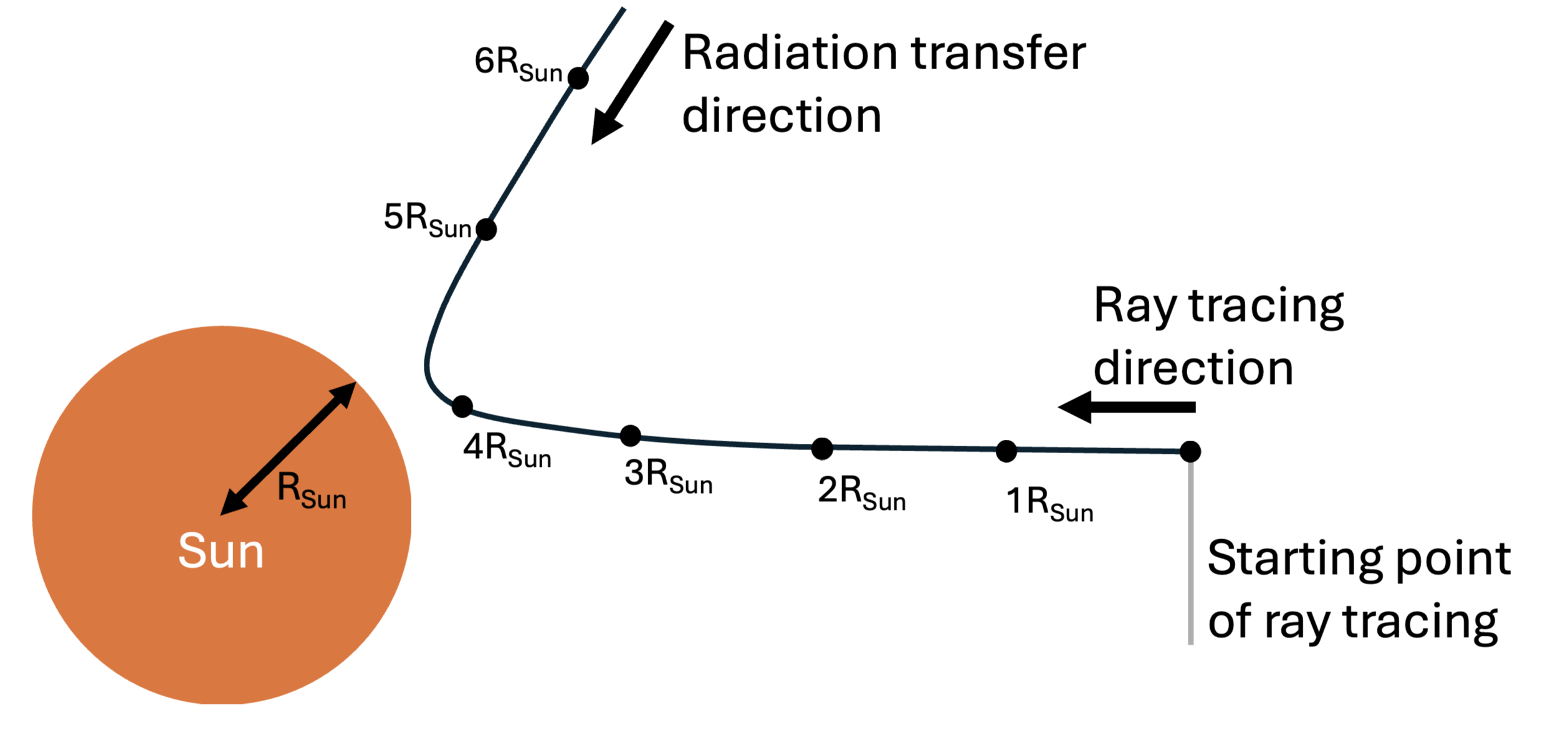}
    \caption{Concept figure for the ray-coordinate convention. The ray coordinate increases along the traced ray toward the Sun (observer to source), which is opposite to the radiative-transfer integration direction (source to observer).}
    \label{fig:concept}
\end{figure}

We perform ray tracing from the observer's perspective by launching rays in the $+\hat{\mathbf{z}}$ direction, with $(x,y)$ defining the image (view) plane. Refraction by density gradients causes rays to deviate from straight lines; an example set of trajectories is shown in Figure~\ref{fig:ray_on_ne}. Along each ray, we estimate the ray-tube cross-sectional change using the step-wise perturbation retracing method described in Section~\ref{sec:ray_tube_scaling}, yielding the local magnification factor $d\ln S/ds$.
As illustrated in Figure~\ref{fig:concept}, for each image pixel the workflow is a two-step process: we first launch a ray from the observer toward the Sun (reverse to the physical radiative-transfer direction) and record the full refracted path and geometric factors. Only after the ray tracing has been completed, we solve the radiative-transfer equation from the end of each traced ray back toward the observer (the actual transfer direction), so the \texttt{GRFF} flux computation waits for the ray-tracing stage to finish.
\begin{figure}[ht!]
    \centering
    \includegraphics[width=0.75\linewidth]{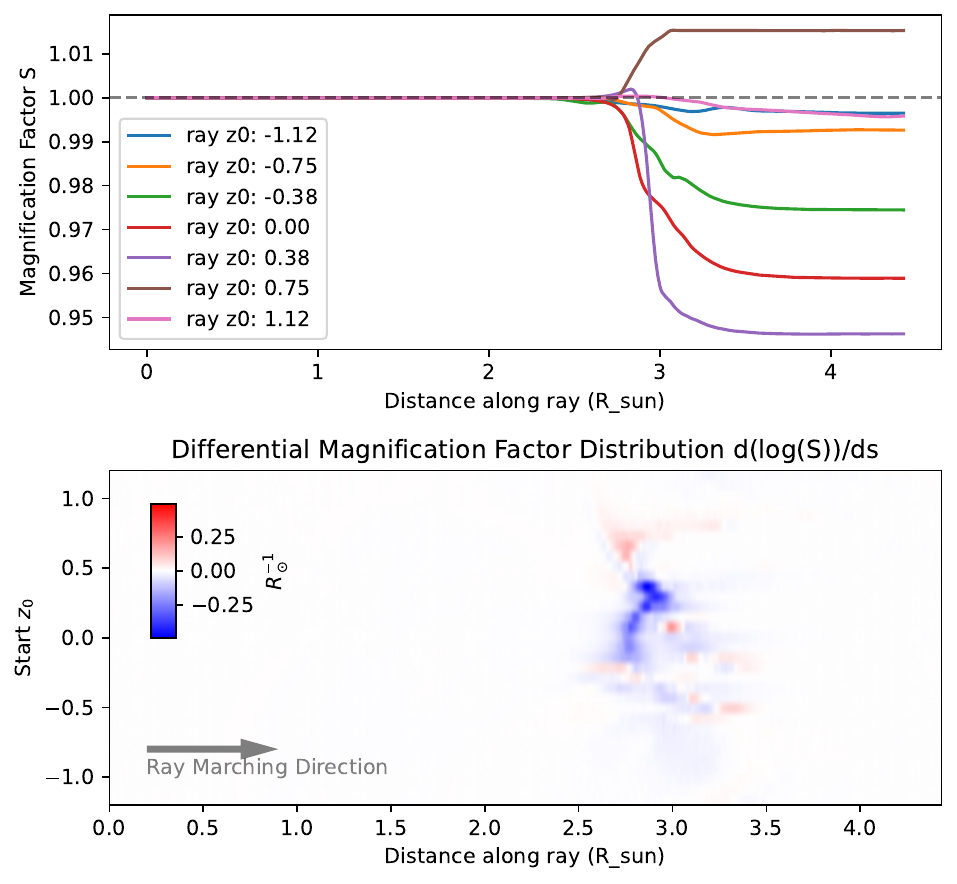}
    \caption{Magnification factor (S) along the ray preceding toward the Sun. Upper panel shows the source area S along the ray path (equavlent to accumulated local magnification factor); values $>1$ indicate defocusing and values $<1$ indicate focusing along the ray, lower panel shows the variation of the magnification factor of rays along the ray path, the rays are lauched from $x=-3.5$, $y=0$, $z=(-1.2, 1.2)$. Figure~\ref{fig:concept} shows the ray-coordinate convention.}
    \label{fig:S_ratio}
\end{figure}
In Figure~\ref{fig:ray_on_ne}, each trajectory is colored by the magnification factor: negative (blue) segments correspond to focusing ($d\ln S/ds<0$) and positive (red) segments correspond to defocusing ($d\ln S/ds>0$). 
Figure~\ref{fig:S_ratio} shows an example of the accumulated magnification factor $S$ and the corresponding magnification factor as a function of distance along the ray.

\subsection{Synthetic imaging results}
\begin{figure}[h!]
    \centering
    \includegraphics[width=0.95\linewidth]{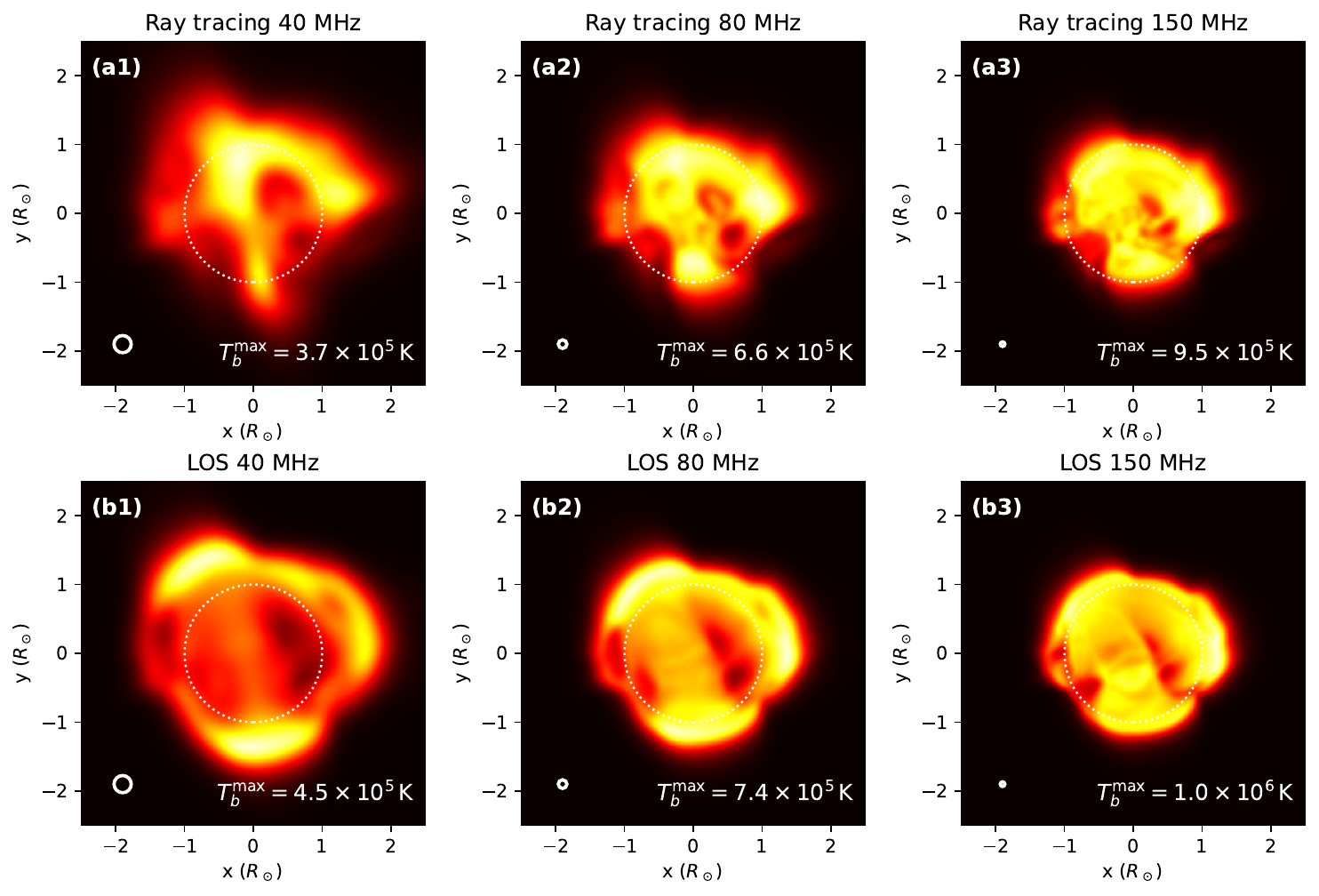}
    \caption{Synthetic quiet-Sun radio images at 40, 80, and 150\,MHz computed from the PSI MAS coronal model with refractive ray tracing and thermal radiative transfer. The model-resolution images are convolved with a representative interferometric beam corresponding to a 15\,km maximum baseline. The top row panels present the result with ray tracing, the bottom row panels present the result without ray tracing. }
    \label{fig:lowbandsynth}
\end{figure}

\begin{figure}[h!]
    \centering
    \includegraphics[width=0.95\linewidth]{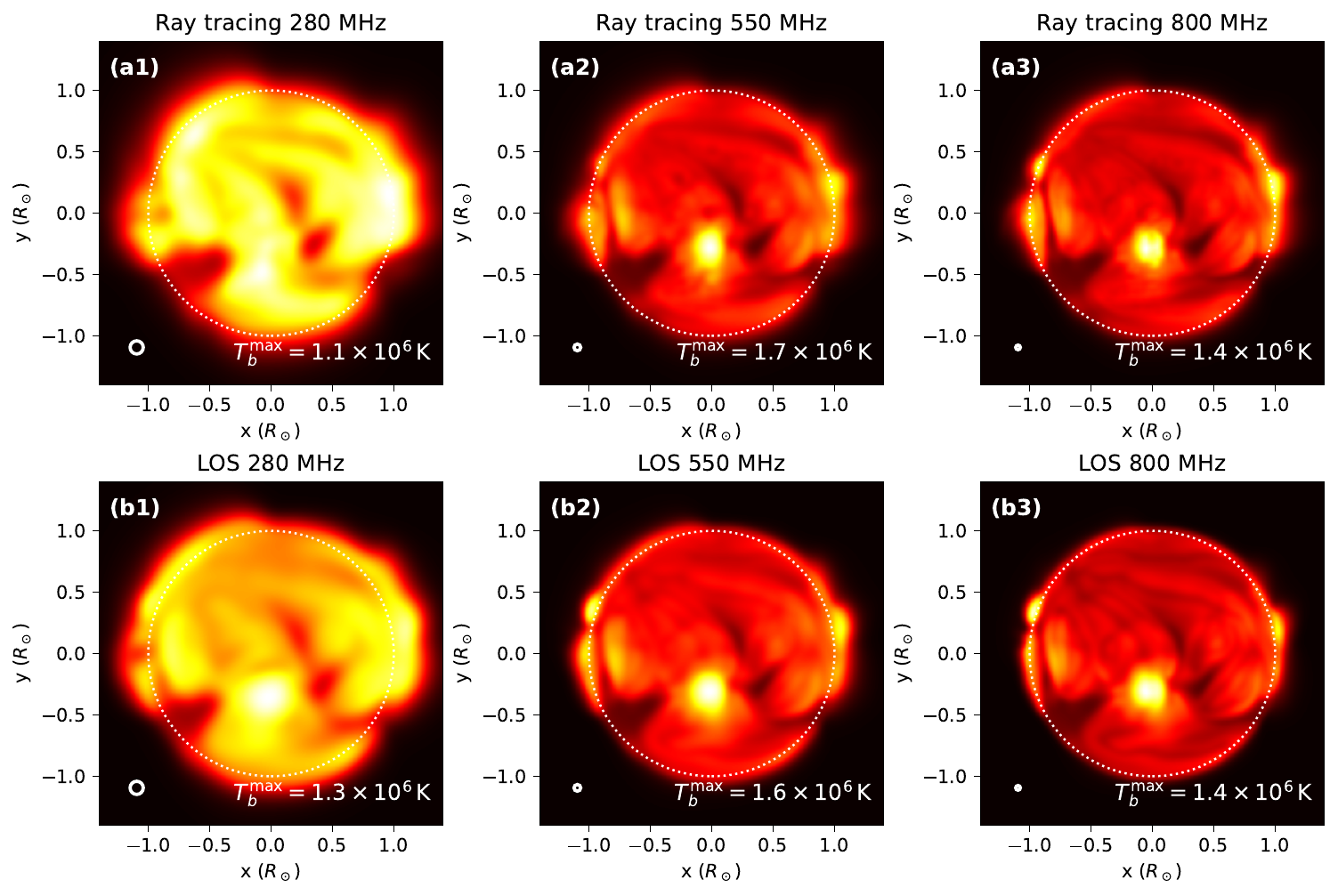}
    \caption{Synthetic quiet-Sun radio images at 280, 550, and 800\,MHz computed from the PSI MAS coronal model using refractive ray tracing and free--free radiative transfer. The intrinsic images are convolved with a representative interferometric beam corresponding to a 5\,km maximum baseline. The dotted circle marks the solar optical limb.}
    \label{fig:highbandsynth}
\end{figure}

After ray tracing, we solve the radiative-transfer equation using the \texttt{GRFF} code \citep{2021ApJ...914...52F}, where the magnification term (Equation~2) was included in the latest release \citep{kuznetsov2021zenodogrff}\footnote{\texttt{GRFF}: \url{https://github.com/kuznetsov-radio/GRFF}; latest Zenodo release: \url{https://doi.org/10.5281/zenodo.19368163}.}. The resulting brightness maps are then formed on the image plane. Figures~\ref{fig:lowbandsynth} and \ref{fig:highbandsynth} show example of the synthetic images at relatively low and high frequencies, respectively.

Figures~\ref{fig:lowbandsynth} and \ref{fig:highbandsynth} compare synthetic images computed with refractive ray tracing (upper panels) to images computed along straight lines of sight (LOS) (lower panels; i.e., neglecting refraction). 
The comparison reveals the following features.

\paragraph{Lower frequencies are more affected by refraction}
The impact of refraction decreases with increasing frequency: the lower the frequency, the larger the difference between the two solutions. 
At 40 and 80~MHz the images with and without ray tracing differ dramatically in both morphology and radial brightness profile. 
In particular, the LOS images exhibit stronger apparent limb brightening.
At higher frequencies, the two approaches show no substantial differences. By 550~MHz and 800~MHz the differences between the ray-traced and straight-line images are not significant at the level visible by eye in the convolved maps.

\paragraph{Spurious limb brightening}
Figure \ref{fig:lowbandsynth} reveals remarkably more limb brightening in the LOS than in the ray-tracing case in the 40 and 80 MHz maps.
The artificially enhanced limb brightening in the LOS case arises because the assumed ray paths sample a different coronal volume than the physically refracted trajectories. The ray-tracing solution shows that rays near the limb bend outward (Figure~\ref{fig:ray_on_ne}), so that the actual propagation paths preferentially pass through plasma with lower density and typically lower temperature (and weaker magnetic field) than the corresponding straight LOS would traverse, as also seen in studies of refractive propagation of solar radio bursts \citep{2011ApJ...734...16T} and in LOFAR quiet-corona imaging with refractive ray tracing \citep{2018A&A...614A..54V}. As a result, the ray-traced paths accumulate a smaller effective emission measure and optical depth, leading to lower free--free emission along the ray and therefore a reduced apparent limb brightening compared with the LOS integration.

\paragraph{Maxium $T_b$ bias}
We also find a systematic tendency for the maximum brightness temperature $T_b$ to be larger in the LOS images than in the corresponding ray-traced solutions, indicating that neglecting refraction correction and the associated ray-tube magnification can bias the peak intensities.

\paragraph{Magnification factor}
Refraction focusing/defocusing changes $S$ along the path, adding a geometric term $d\ln S/ds$ to the radiative-transfer equation (Section~\ref{sec:ray_tube_scaling}). By flux conservation, a shrinking tube increases the observed intensity, while an expanding tube decreases it.

\begin{figure}
    \centering
    \includegraphics[width=0.99\linewidth]{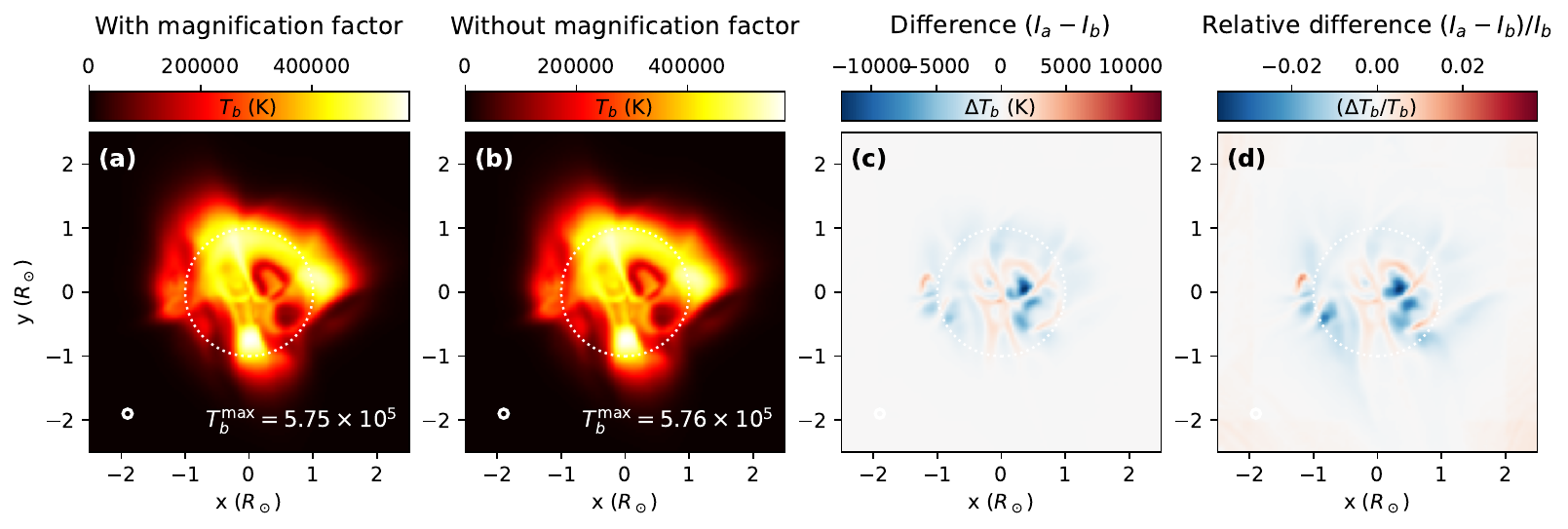}
    \caption{Effect of the ray-tube magnification term in the radiative-transfer calculation. Shown are synthetic quiet-Sun brightness maps at 60~MHz computed with refractive ray tracing from the same coronal model, comparing solutions that include and omit the geometric magnification factor proportional to $d\ln S/ds$.}
    \label{fig:mf}
\end{figure}

For the density model we use in this case, the magnitude of this correction is modest. Figure~\ref{fig:mf} shows that including the magnification term changes the brightness temperature by less than $\sim 4\%$ and does not introduce visible morphological differences in the synthetic images. Consistent with this, Figure~\ref{fig:S_ratio} indicates that along traced rays $d\ln S/ds$ remains below $\sim 0.4 \,\rm R_\odot^{-1}$, and the cumulative change of the tube area is small: the maximum integrated factor $S/S_0$ differs from unity by only $\sim 6\%$.  We therefore conclude that the magnification factor provides a formally flux-conserving correction but is a second-order effect in our simulations, indicating that the dominant propagation effect is ray bending rather than ray focusing. The magnification term does not change the overall image morphology.

\subsection{Comparison with observations}

%To further validate the model, we compare it with radio observation from OVRO-LWA.

\subsubsection{Brightness temperature spectrum}

The near-center ($<0.5R_\odot$) quiet-Sun brightness temperature spectrum can be derived from the synthetic images. For each frequency, we measure the mean $T_b$ within the model-predicted solar disk and compare the resulting spectrum with historical quiet-Sun measurements reported in the literature. 

\begin{figure}[h!]
    \centering
    \includegraphics[width=0.81\linewidth]{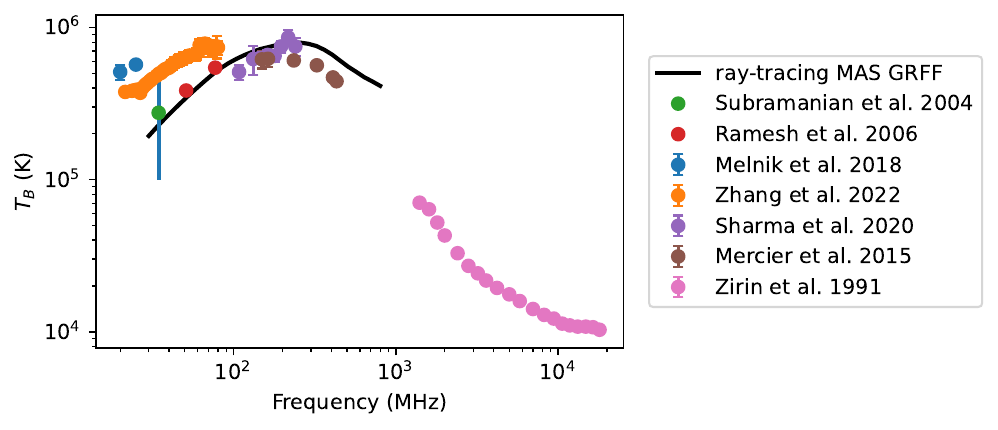}
    \caption{Quiet-Sun brightness temperature spectrum. The solid curve shows the modeled $<0.5R_\odot$ averaged $T_b$ derived from our synthetic images over 40--800~MHz. Symbols show representative measurements from previous quiet-Sun observations in the literature (e.g., \citealt{2004A&A...426..329S,2006ApJ...648..707R,2018SoPh..293...97M,2022ApJ...932...17Z}).}
    \label{fig:Tb_spec}
\end{figure}

Figure~\ref{fig:Tb_spec} shows the modeled $T_b(\nu)$ over 40--800~MHz together with representative observational constraints from previous studies (e.g., \citealt{2004A&A...426..329S,2006ApJ...648..707R,2018SoPh..293...97M,2022ApJ...932...17Z}). The modeled brightness temperatures are consistent with the observed range across the band: the spectrum decreases from the lowest frequencies toward a few hundred MHz as the effective formation height moves downward and the emission becomes progressively less optically thick, and then flattens at the highest frequencies where the quiet-Sun disk approaches the optically thin regime. The remaining differences are at the level expected given the snapshot nature of the coronal model, uncertainties in absolute flux calibration among different instruments, and the fact that scattering and fine-scale density structure are not included in the present calculations.

\subsubsection{Solar images compared with OVRO LWA}

\begin{figure}[h!]
    \centering
    \includegraphics[height=0.68\linewidth]{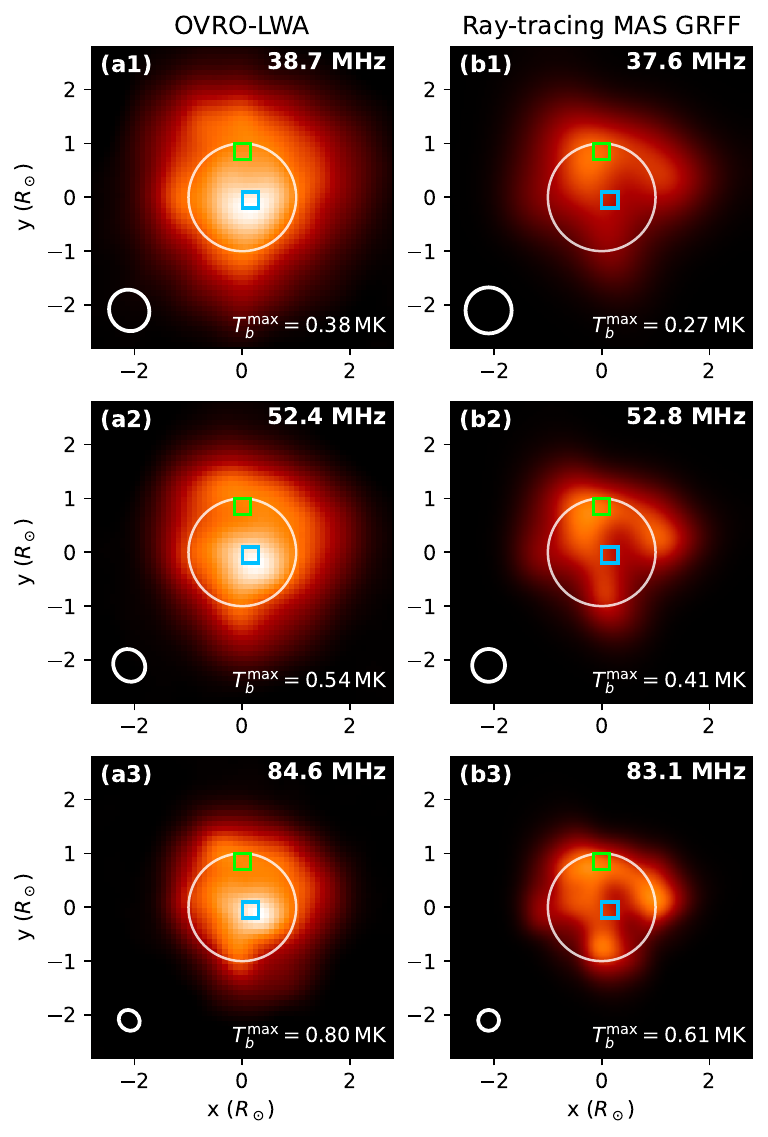}
    \includegraphics[height=0.62\linewidth]{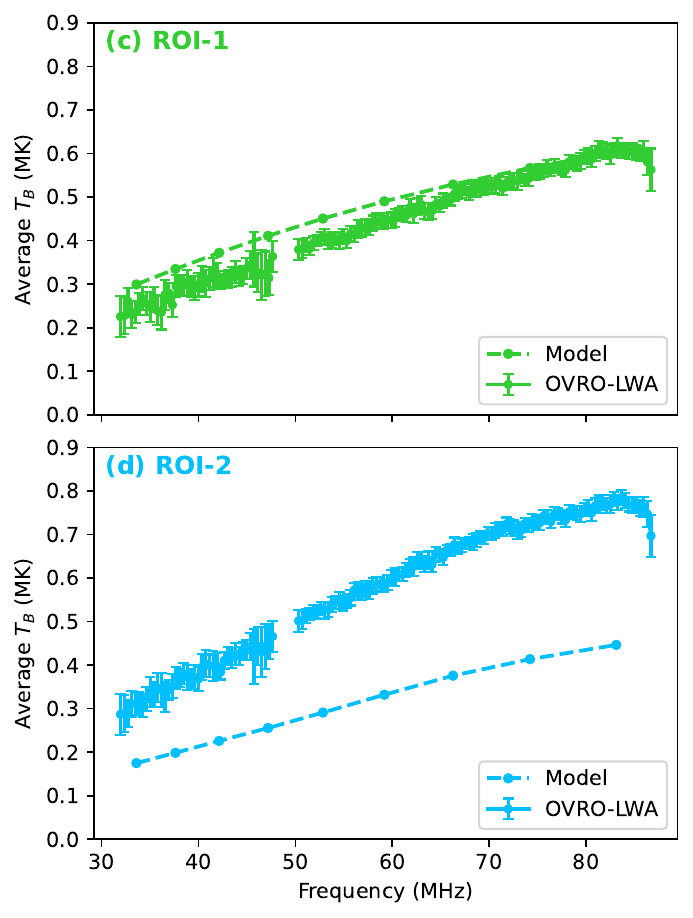}
    \caption{
Comparison between observed and modeled Sun brightness at decameter wavelengths.
Panels (a1--a3) show OVRO--LWA quiet-Sun images from 2025-06-08T20:07:03~UT at 38.7, 52.4, and 84.6~MHz.
Panels (b1--b3) show the corresponding synthetic images computed with refractive ray tracing and GRFF radiative transfer using the PSI/MAS coronal model for Carrington rotation 2298, rotated to align the coronal configuration to 2025-06-08T20:00:00~UT.
Two regions of interest (ROIs) are marked in all images: ROI--1 (green square; quiet region) and ROI--2 (blue square; near an active-region vicinity).
Panels (c) and (d) show the mean brightness temperature spectra (average $T_B$) within ROI--1 and ROI--2, respectively, comparing the OVRO--LWA measurements with the model predictions across $\sim$30--85~MHz.
}
    \label{fig:ovro_mas_roi}
\end{figure}

We present a direct comparison between OVRO--LWA quiet-Sun images observed on 2025-06-08T20:07:03~UT and synthetic images produced by the refractive ray-tracing MAS--GRFF pipeline using the Carrington rotation 2298 MAS solution rotated to the observing epoch (Figure~\ref{fig:ovro_mas_roi}). The overall morphology and brightness level are consistent: the modeled thermal emission (free--free, with gyro-resonance where applicable) reproduces the dominant background quiet-Sun component at decameter wavelengths, supporting the use of the MAS plasma parameters as a physically motivated baseline for incoherent emission. 

To quantify local spectral behavior, we define two regions of interest (ROIs) shown in Figure~\ref{fig:ovro_mas_roi}. ROI--1 (green) is placed in a relatively quiet region, while ROI--2 (blue) samples a location closer to an active-region vicinity near disk center. The resulting brightness-temperature spectra are shown in panels (c) and (d). In the quiet ROI--1, the observed spectrum follows the modeled trend closely across $\sim$30--85~MHz, indicating that thermal emission and refraction captured by the MAS--GRFF framework can explain the local brightness and its frequency dependence in quiet areas. In contrast, ROI--2 exhibits a clear mismatch: the observed $T_B$ is systematically higher than the model and the spectral shape deviates, suggesting that additional physics and/or model limitations become important near active-region environments.

\begin{figure}[h!]
    \centering
    \includegraphics[width=0.95\linewidth]{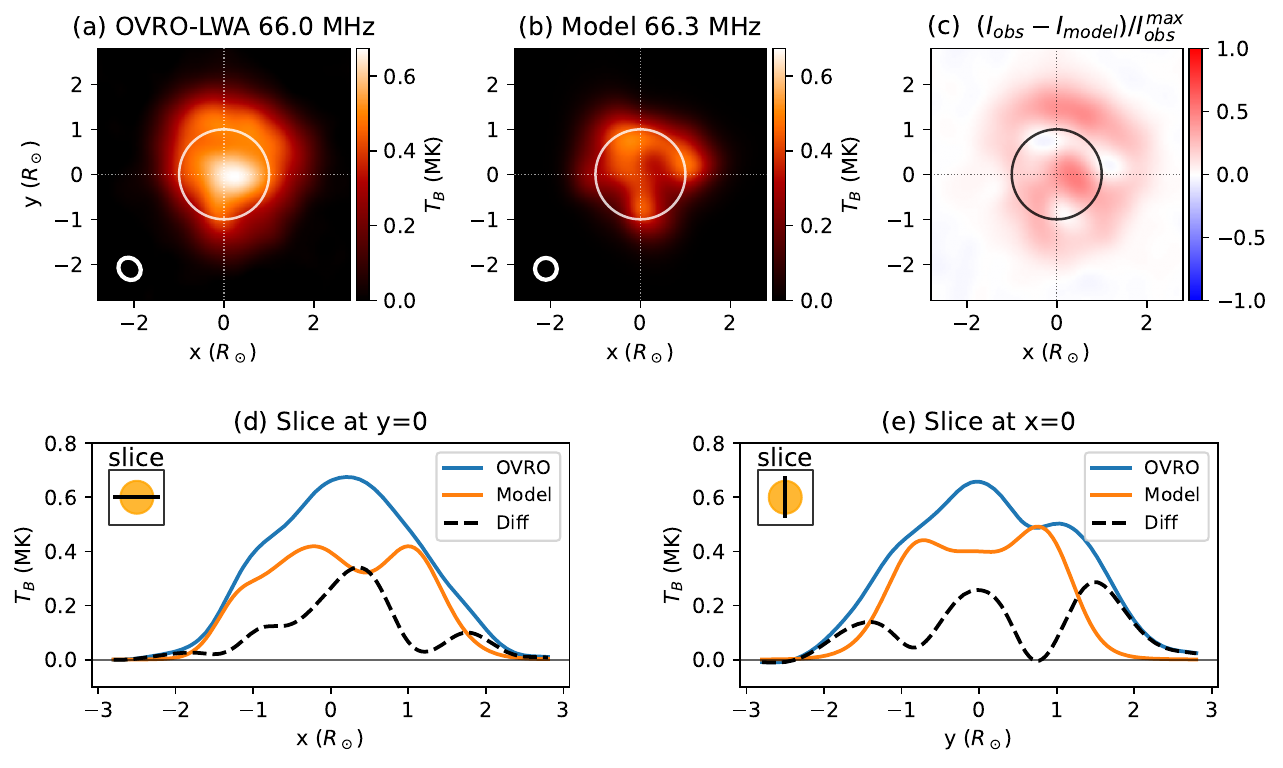}
    \caption{
Example image-level comparison between OVRO--LWA observations and the ray-tracing MAS--GRFF model at $\sim$66~MHz.
(a) OVRO--LWA quiet-Sun brightness temperature map at 66.0~MHz.
(b) Corresponding synthetic brightness temperature map from the model at 66.3~MHz.
(c) Normalized difference map, $(I_{\rm obs}-I_{\rm model})/I_{\rm obs}^{\max}$, shown with a diverging color scale.
The black circle marks the solar disk, and the dashed lines indicate the $x=0$ and $y=0$ axes used for 1D cuts.
(d) Horizontal slice at $y=0$ comparing the observed (blue) and modeled (orange) $T_B$ profiles; the dashed black curve shows their difference.
(e) Vertical slice at $x=0$ with the same format.
}

    \label{fig:slicediff}
\end{figure}

Figure~\ref{fig:slicediff} further highlights where the model departs from the data using an example at $\sim$66~MHz. The normalized difference map shows the largest residuals concentrated (1) near the active-region vicinity, consistent with the ROI--2 behavior discussed above, and (2) toward the limb where the observed emission extends to larger apparent radii than predicted by the model. The latter is naturally explained by radio-wave scattering from coronal density fluctuations, which can broaden the apparent source and enhance off-limb extension; such scattering is not included in the current MAS--GRFF forward modeling and is expected to be increasingly important at these frequencies (see, e.g., \citet{1994ApJ...426..774B,Kontar2019}). Together, Figures~\ref{fig:ovro_mas_roi} and \ref{fig:slicediff} indicate that while thermal emission plus refraction can reproduce quiet-region spectra and gross morphology, accurate modeling of active-region neighborhoods and limb extension likely requires improved coronal thermodynamics and explicit treatment of scattering and/or weak coherent emission.

There are several factors that may contribute to the discrepancy observed at ROI 2. 
First, due to its modest size, the global MHD solution may not resolve the active-region thermodynamic structure accurately; in particular, an underestimated coronal electron temperature $T_e$ and/or density $n_e$ in the low corona would directly reduce the modeled free--free emissivity and optical depth, lowering $T_B$ relative to observations. 
Second, weak coherent contributions (e.g., low-level plasma radiation or electron-cyclotron maser emission) may occur intermittently or persistently near active regions \citep{2024NatAs...8...50Y}, 
elevating the observed brightness above our prediction based on a thermal emission model even during otherwise quiet conditions. 
Additional contributors include (i) unresolved fine-scale density structuring not captured by the MAS grid (affecting both emissivity and propagation), (ii) mismatch of temporal evolution between the MAS model, which is based on an input synoptic magnetogram from an entire Carrington rotation of 27 days, and the actual observation, which is based on a single 10-s snapshot.

\section{Discussion and Summary}
\label{sec:discussion_summary}

We have developed a forward-modeling framework to synthesize solar radio images at sub-GHz frequencies. This framework combines refractive ray tracing using a 3D MHD coronal model with radiative transfer, including a local ray-tube area (magnification) correction. In this approach, refraction in the inhomogeneous corona bends rays and changes the cross-sectional area of a ray bundle $S(s)$, which introduces an additional geometric term $\mathrm{d}\ln S/\mathrm{d}s$ in the transfer equation to ensure flux conservation under focusing/defocusing.

Our synthetic imaging experiments reinforce that refraction is the dominant propagation effect shaping thermal radio Sun morphology. Comparisons between ray-traced images and straight LOS integrations show that refraction has the largest impact at low frequencies, where the two solutions differ substantially in both morphology and brightness temperature slices. As the frequency increases, the two approaches converge, as expected, because refractive bending weakens as the refractive index approaches unity. By $\sim$280~MHz the ray-traced and LOS maps are already similar, although residual differences remain discernible; above $\sim$550~MHz the differences become small, with only subtle differences persisting.
These trends connect naturally to prior thermal Sun forward modeling. For example, \citet{2020ApJ...903..126S} compared MWA quiet-Sun observations with FORWARD-based thermal bremsstrahlung maps driven by a MAS coronal model and found systematic discrepancies (e.g., limb-related differences) attributable to propagation effects. Synthetic LOS maps display spurious limb brightening at low frequencies because the assumed paths traverse denser/hotter coronal volumes, located lower in the solar corona than the physically refracted trajectories. In our calculations, ray tracing reduces this bias: rays near the limb bend outward and preferentially sample lower-density plasma, lowering the effective emission measure and optical depth along the true paths and weakening the apparent limb brightening relative to LOS integrations. Thus, explicitly modeling refraction provides an improved, physically grounded route to reconcile thermal forward models with observed disk morphologies at thermal radio frequencies.

We implemented a step-wise perturbation re-tracing algorithm to estimate the evolution of the local magnification factor along each ray. By launching perturbed rays in the local transverse plane and measuring the resulting separation vectors after each integration step, we compute $S(s)$ and incorporate it self-consistently in the transfer solver. For the tested thermal radio solar configuration, including this magnification term produces measurable but not dominant changes: it modifies $T_B$ at the few-percent level and does not introduce visible morphological changes in the synthetic images. This indicates that, for typical thermal radio solar conditions, refraction primarily affects the image through ray bending (that defines, which coronal volumes are sampled), while the additional focusing/defocusing correction is a smaller effect.

We also compared our model with OVRO--LWA imaging at decameter wavelengths (Figures~\ref{fig:ovro_mas_roi} and \ref{fig:slicediff}). The overall morphology and brightness level are consistent, indicating that the combination of free--free and gyroresonance emission represent the dominant thermal radio solar background when driven by the MAS model. The agreement is best in a quiet region (ROI 1), where the observed $T_B(\nu)$ follows the modeled trend across $\sim$30--85~MHz (Figure~\ref{fig:ovro_mas_roi}c). In contrast, the spectrum shows a larger discrepancy in the vicinity of an active region close to the disk center (ROI--2; Figure~\ref{fig:ovro_mas_roi}d): the observations are systematically brighter than the model and the spectral shape deviates, suggesting that additional physics and/or model limitations become important near active regions.

Several factors may contribute to the ROI--2 discrepancy. First, the global coronal model may not represent active-region thermodynamics with sufficient fidelity; for example, an underestimated electron temperature $T_e$ (and/or density $n_e$) in the low corona would directly reduce the modeled free--free emissivity and optical depth, lowering $T_B$ relative to the data. Second, weak coherent contributions may elevate the observed brightness even during nominally quiet conditions near active regions, e.g., low-level plasma emission or electron-cyclotron maser emission not captured in a purely incoherent forward model. Additional contributors include unresolved fine-scale density structuring (which affects both emissivity and propagation), and temporal evolution between the MAS snapshot and the observation time.

Figure~\ref{fig:slicediff} further shows that the largest residuals occur near the active-region vicinity and toward the limb and off-limb. The active-region discrepancy is consistent with the behavior of ROI 2 discussed above. The limb-related difference is characterized by the observed emission extending to larger apparent radial distances than the model prediction. 
One possible explanation is radio-wave scattering from coronal density fluctuations, which broadens the apparent source and redistributes intensity outward; such scattering is not included in the present forward modeling and is expected to be important at these frequencies (e.g., \citealt{1994ApJ...426..774B,Kontar2019}). Incorporating scattering is therefore a clear priority for improving the realism of the synthetic images, particularly for reproducing the off-limb extension and the detailed radial profiles.

The combination of well-calibrated interferometric imaging and physics-based forward modeling offers a path toward constraining the coronal background more directly. If the dominant thermal emission and propagation effects (refraction and scattering) are captured with sufficient fidelity, then discrepancies between observed and synthesized brightness maps can be used as quantitative constraints on the underlying coronal parameters $(T_e,\,n_e,\,\mathbf{B})$. In this perspective, multi-frequency imaging can be used not only to validate coronal models but also to potentially improve MAS model through iterative parameter tuning or assimilation, ultimately advancing our understanding of the coronal background and its spatial variability.

\keywords{}

\begin{acknowledgments}
P.Z. acknowledge google cloud for computing resource support. This material is based upon work supported by the Google Cloud Research Credits program with the award GCP19980904. The OVRO-LWA expansion project was supported by NSF under grant AST-1828784. OVRO-LWA solar operations are supported by NSF grant AGS-2436999 to the New Jersey Institute of Technology. %We acknowledge Alexey Kuznetsov for his foundational work on GRFF and the development of the associated algorithms.
C.D. acknowledges support from the NASA Living With a Star Strategic Capabilities program (award 80NSSC22K0893).
\end{acknowledgments}

%\begin{contribution}
%\end{contribution}

\facilities{OVRO-LWA}

\bibliography{cite}

@ARTICLE{2023A&A...670C...5A,
       author = {{Alissandrakis}, C.~E. and {Bastian}, T.~S. and {Nindos}, A.},
        title = "{A first look at the submillimeter Sun with ALMA (Corrigendum)}",
      journal = {\aap},
         year = 2023,
        month = feb,
       volume = {670},
          eid = {C5},
        pages = {C5},
          doi = {10.1051/0004-6361/202243774e},
       adsurl = {https://ui.adsabs.harvard.edu/abs/2023A&A...670C...5A}
}

@ARTICLE{Kontar2019,
       author = {{Kontar}, Eduard P. and {Chen}, Xingyao and {Chrysaphi}, Nicolina and {Jeffrey}, Natasha L.~S. and {Emslie}, A. Gordon and {Krupar}, Vratislav and {Maksimovic}, Milan and {Gordovskyy}, Mykola and {Browning}, Philippa K.},
        title = "{Anisotropic Radio-wave Scattering and the Interpretation of Solar Radio Emission Observations}",
      journal = {\apj},
         year = 2019,
        month = oct,
       volume = {884},
       number = {2},
          eid = {122},
        pages = {122},
          doi = {10.3847/1538-4357/ab40bb},
archivePrefix = {arXiv},
       eprint = {1909.00340},
 primaryClass = {astro-ph.SR},
       adsurl = {https://ui.adsabs.harvard.edu/abs/2019ApJ...884..122K}
}

@article{2022ApJ...932...17Z,
  author  = {Zhang, PeiJin and Zucca, Pietro and Kozarev, Kamen and Carley, Eoin and Wang, ChuanBing and Franzen, Thomas and Dabrowski, Bartosz and Krankowski, Andrzej and Magdaleni{\'c}, Jasmina and Vocks, Christian},
  title   = {Imaging of the Quiet Sun in the Frequency Range of 20--80 MHz},
  journal = {The Astrophysical Journal},
  year    = {2022},
  volume  = {932},
  eid     = {17},
  doi     = {10.3847/1538-4357/ac6b37},
  adsurl  = {https://ui.adsabs.harvard.edu/abs/2022ApJ...932...17Z/abstract}
}

@article{1991ApJ...370..779Z,
  author  = {Zirin, H. and Baumert, B. M. and Hurford, G. J.},
  title   = {The Microwave Brightness Temperature Spectrum of the Quiet Sun},
  journal = {The Astrophysical Journal},
  year    = {1991},
  volume  = {370},
  pages   = {779},
  doi     = {10.1086/169861},
  adsurl  = {https://ui.adsabs.harvard.edu/abs/1991ApJ...370..779Z/abstract}
}

@article{2004A&A...426..329S,
  author  = {Subramanian, K. R.},
  title   = {Brightness temperature and size of the quiet Sun at 34.5 MHz},
  journal = {Astronomy \& Astrophysics},
  year    = {2004},
  volume  = {426},
  pages   = {329--331},
  doi     = {10.1051/0004-6361:20047120}
}

@article{2006ApJ...648..707R,
  author  = {Ramesh, R. and Nataraj, H. S. and Kathiravan, C. and Sastry, Ch. V.},
  title   = {The Equatorial Background Solar Corona during Solar Minimum},
  journal = {The Astrophysical Journal},
  year    = {2006},
  volume  = {648},
  pages   = {707--711},
  doi     = {10.1086/505677}
}

@article{2015A&A...583A.101M,
  author  = {Mercier, C. and Chambe, G.},
  title   = {Electron density and temperature in the solar corona from multifrequency radio imaging},
  journal = {Astronomy \& Astrophysics},
  year    = {2015},
  volume  = {583},
  eid     = {A101},
  doi     = {10.1051/0004-6361/201425540}
}

@article{2018SoPh..293...97M,
  author  = {Melnik, V. N. and Shepelev, V. A. and Poedts, S. and Dorovskyy, V. V. and Brazhenko, A. I. and Rucker, H. O.},
  title   = {Interferometric Observations of the Quiet Sun at 20 and 25 MHz in May 2014},
  journal = {Solar Physics},
  year    = {2018},
  volume  = {293},
  eid     = {97},
  doi     = {10.1007/s11207-018-1316-3},
  adsurl  = {https://ui.adsabs.harvard.edu/abs/2018SoPh..293...97M/abstract}
}

@article{2020ApJ...903..126S,
  author  = {Sharma, Rohit and Oberoi, Divya},
  title   = {Propagation Effects in Quiet Sun Observations at Meter Wavelengths},
  journal = {The Astrophysical Journal},
  year    = {2020},
  volume  = {903},
  eid     = {126},
  doi     = {10.3847/1538-4357/abb949},
  adsurl  = {https://ui.adsabs.harvard.edu/abs/2020ApJ...903..126S/abstract}
}

@ARTICLE{2021ApJ...914...52F,
       author = {{Fleishman}, Gregory D. and {Kuznetsov}, Alexey A. and {Landi}, Enrico},
        title = "{Gyroresonance and Free-Free Radio Emissions from Multithermal Multicomponent Plasma}",
      journal = {\apj},
         year = 2021,
        month = jun,
       volume = {914},
       number = {1},
        pages = {52},
          doi = {10.3847/1538-4357/abf92c},
       adsurl = {https://ui.adsabs.harvard.edu/abs/2021ApJ...914...52F}
}

@MISC{kuznetsov2021zenodogrff,
       author = {{Kuznetsov}, Alexey A. and {Fleishman}, Gregory D. and {Landi}, Enrico},
        title = "{Codes for computing the solar gyroresonance and free-free radio emissions, v1.0.1}",
         year = 2026,
    publisher = {Zenodo},
          doi = {10.5281/zenodo.19368163},
          url = {https://zenodo.org/records/19368163}
}

@INPROCEEDINGS{1996AIPC..382..104M,
       author = {{Miki{\'c}}, Zoran and {Linker}, Jon A.},
        title = "{The large-scale structure of the solar corona and inner heliosphere}",
    booktitle = {American Institute of Physics Conference Series},
         year = 1996,
       series = {American Institute of Physics Conference Series},
       volume = {382},
        pages = {104-107},
          doi = {10.1063/1.51370},
       adsurl = {https://ui.adsabs.harvard.edu/abs/1996AIPC..382..104M}
}

@ARTICLE{1994ApJ...426..774B,
       author = {{Bastian}, T.~S.},
        title = "{Angular Scattering of Solar Radio Emission by Coronal Turbulence}",
      journal = {\apj},
         year = 1994,
        month = may,
       volume = {426},
        pages = {774},
          doi = {10.1086/174114},
       adsurl = {https://ui.adsabs.harvard.edu/abs/1994ApJ...426..774B}
}

@ARTICLE{1985ARA&A..23..169D,
       author = {{Dulk}, George A.},
        title = "{Radio emission from the Sun and stars}",
      journal = {\araa},
         year = 1985,
       volume = {23},
        pages = {169--224},
          doi = {10.1146/annurev.aa.23.090185.001125},
       adsurl = {https://ui.adsabs.harvard.edu/abs/1985ARA%26A..23..169D}
}

@ARTICLE{2011SoPh..273..309S,
       author = {{Shibasaki}, Kiyoto and {Alissandrakis}, C.~E. and {Pohjolainen}, S.},
        title = "{Radio Emission of the Quiet Sun and Active Regions}",
      journal = {\solphys},
         year = 2011,
       volume = {273},
        pages = {309--337},
          doi = {10.1007/s11207-011-9788-4},
       adsurl = {https://ui.adsabs.harvard.edu/abs/2011SoPh..273..309S}
}

@ARTICLE{2011ApJ...734...16T,
       author = {{Thejappa}, G. and {MacDowall}, R.~J. and {Gopalswamy}, N.},
        title = "{Effects of Refraction on Angles and Times of Arrival of Solar Radio Bursts}",
      journal = {\apj},
         year = 2011,
        month = jun,
       volume = {734},
       number = {1},
          eid = {16},
        pages = {16},
          doi = {10.1088/0004-637X/734/1/16},
       adsurl = {https://ui.adsabs.harvard.edu/abs/2011ApJ...734...16T}
}

@ARTICLE{2018A&A...614A..54V,
       author = {{Vocks}, C. and {Mann}, G. and {Breitling}, F. and {Bisi}, M.~M. and {D{\k{a}}browski}, B. and {Fallows}, R. and {Gallagher}, P.~T. and {Krankowski}, A. and {Magdaleni{\'c}}, J. and {Marqu{\'e}}, C. and {Morosan}, D. and {Rucker}, H.},
        title = "{LOFAR observations of the quiet solar corona}",
      journal = {\aap},
         year = 2018,
        month = jun,
       volume = {614},
          eid = {A54},
        pages = {A54},
          doi = {10.1051/0004-6361/201630067},
archivePrefix = {arXiv},
       eprint = {1803.00453},
 primaryClass = {astro-ph.SR},
       adsurl = {https://ui.adsabs.harvard.edu/abs/2018A&A...614A..54V}
}

@ARTICLE{1971A&A....10..362S,
       author = {{Steinberg}, J.~L. and {Aubier-Giraud}, M. and {Leblanc}, Y. and {Boischot}, A.},
        title = "{Coronal Scattering, Absorption and Refraction of Solar Radiobursts}",
      journal = {\aap},
         year = 1971,
        month = feb,
       volume = {10},
        pages = {362--365},
       adsurl = {https://ui.adsabs.harvard.edu/abs/1971A&A....10..362S}
}

@ARTICLE{2026ApJ...999..237M,
       author = {{Mondal}, Surajit and {Shaik}, Shaheda Begum and {Howard}, Russell A. and {Zhang}, Peijin and {Chen}, Bin and {Chen}, Xingyao and {Yu}, Sijie and {Gary}, Dale and {Anderson}, Marin M. and {Bowman}, Judd D. and {Byrne}, Ruby and {Catha}, Morgan and {Chhabra}, Sherry and {D'Addario}, Larry and {Davis}, Ivey and {Dowell}, Jayce and {Hallinan}, Gregg and {Harnach}, Charlie and {Hellbourg}, Greg and {Hickish}, Jack and {Hobbs}, Rick and {Hodge}, David and {Hodges}, Mark and {Huang}, Yuping and {Isella}, Andrea and {Jacobs}, Daniel C. and {Kemby}, Ghislain and {Klinefelter}, John T. and {Kolopanis}, Matthew and {Kosogorov}, Nikita and {Lamb}, James and {Law}, Casey and {Mahesh}, Nivedita and {O'Donnell}, Brian and {Plant}, Kathryn and {Posner}, Corey and {Powell}, Travis and {Prayag}, Vinand and {Rizo}, Andres and {Romero-Wolf}, Andrew and {Shi}, Jun and {Taylor}, Greg and {Trim}, Jordan and {Virgin}, Mike and {Vydula}, Akshatha and {Weinreb}, Sandy and {White}, Scott and {Woody}, David and {Zentmeyer}, Thomas},
        title = "{Estimating Electron Densities in the Middle Solar Corona Using White-light and Radio Observations}",
      journal = {\apj},
         year = 2026,
        month = mar,
       volume = {999},
       number = {2},
          eid = {237},
        pages = {237},
          doi = {10.3847/1538-4357/ae4353},
archivePrefix = {arXiv},
       eprint = {2602.09819},
 primaryClass = {astro-ph.SR},
       adsurl = {https://ui.adsabs.harvard.edu/abs/2026ApJ...999..237M}
}

@ARTICLE{2009ApJ...690..902L,
       author = {{Lionello}, Roberto and {Linker}, Jon A. and {Miki{\'c}}, Zoran},
        title = "{Multispectral Emission of the Sun During the First Whole Sun Month: Magnetohydrodynamic Simulations}",
      journal = {\apj},
         year = 2009,
        month = jan,
       volume = {690},
       number = {1},
        pages = {902-912},
          doi = {10.1088/0004-637X/690/1/902},
       adsurl = {https://ui.adsabs.harvard.edu/abs/2009ApJ...690..902L}
}

@ARTICLE{2024NatAs...8...50Y,
       author = {{Yu}, Sijie and {Chen}, Bin and {Sharma}, Rohit and {Bastian}, Timothy S. and {Mondal}, Surajit and {Gary}, Dale E. and {Luo}, Yingjie and {Battaglia}, Marina},
        title = "{Detection of long-lasting aurora-like radio emission above a sunspot}",
      journal = {Nature Astronomy},
         year = 2024,
        month = jan,
       volume = {8},
       number = {1},
        pages = {50--59},
          doi = {10.1038/s41550-023-02122-6},
       adsurl = {https://ui.adsabs.harvard.edu/abs/2024NatAs...8...50Y}
}
\bibliographystyle{aasjournalv7}

\end{document}